\documentstyle[aps,epsf,twocolumn]{revtex}
\newcommand{\be}{\begin{eqnarray}}
\newcommand{\ee}{\end{eqnarray}}
\begin{document}
\draft
\title{\bf  The Pion-Nucleon Sigma Term and the Goldberger-Treiman
Discrepancy}
\author{{\bf James V. Steele}$^1$, {\bf Hidenaga Yamagishi}
$^2$ and {\bf Ismail Zahed}$^1$}
\address{$^1$Department of Physics, SUNY, Stony Brook, New York 11794,
USA;\\ $^2$4 Chome 11-16-502, Shimomeguro, Meguro, Tokyo, Japan. 153}
\date{\today}
\maketitle

\begin{abstract}
The pion-nucleon sigma term is shown to be equal to the
Goldberger-Treiman discrepancy at tree level. Its value estimated
this way is very sensitive to the pion-nucleon coupling constant
$g_{\pi NN}$. This relation, when combined with the pion-nucleon
S-wave scattering lengths, yields a new determination of $g_{\pi NN}$
at tree level. The results of a one-loop analysis are also summarized
determining an allowed range for the induced pseudoscalar coupling
constant $g_P$.
\end{abstract}
\narrowtext

Pion-nucleon interactions have been extensively investigated using
dispersion relations and chiral symmetry. Most of the studies using
chiral symmetry have relied on unphysical limits such as the soft pion
limit \cite{ADLER} or the chiral limit \cite{DASHEN}. A typical
example is the pion-nucleon sigma term \cite{SIGMA}, the fraction of
the nucleon mass due to the explicit breaking of chiral $SU(2)\times
SU(2)$. The scattering amplitude is analytically continued to the
unphysical Cheng-Dashen point \cite{CHENG}, and chiral perturbation
theory is applied.

An important exception to the above is Weinberg's formula for
pion-nucleon scattering \cite{WEINBERG}, which yields the
Tomozawa-Weinberg relations for the S-wave scattering lengths on shell
\cite{TOMOZAWA}. Recently, we have been able to extend this result to
processes involving an arbitrary number of on-shell pions and nucleons
\cite{YAMZAH1,YAMZAH2}. In this way, the pion-nucleon sigma term can
be directly assessed. In particular, we find that at tree level the
pion-nucleon sigma term is simply given by the Goldberger-Treiman
discrepancy.  The purpose of this letter is to give a derivation of
this result, and discuss some of its quantitative aspects. We also
review Weinberg's formula in light of our result, and briefly discuss
the effects of one-loop corrections.

The approach discussed in \cite{YAMZAH1,YAMZAH2} requires an extended
S matrix analysis for a concise quantum formulation that enforces
both chiral symmetry and unitarity. However, since we are primarily
interested here in a tree level result, we will use an equivalent but
shorter route in terms of effective Lagrangians with some supplemental
rules following from the complete analysis \cite{YAMZAH1,YAMZAH2}.

For the $SU(2)\times SU(2)$ symmetric part, we take the standard effective
Lagrangian
\be
{\cal L}_1 = &&\frac{f_{\pi}^2}4 \,\, {\rm Tr}
\left( \partial_{\mu} U\,\, \partial^{\mu} U^{\dagger}
\right)\nonumber\\&&
+\, \overline{\bf \Psi} i\rlap/\partial {\bf \Psi}
-m_{\rm inv} \, \left( \overline{\bf \Psi}_R U {\bf \Psi}_L
+ \overline{\bf \Psi}_L U^{\dagger} {\bf \Psi}_R \right)\nonumber\\
&& +\frac i2 (g_A-1)
   \overline{\bf \Psi}_R (\rlap/\partial U )
U^{\dagger} {\bf \Psi}_R
\nonumber\\
&&-\frac i2 (g_A-1) \,\overline{\bf \Psi}_L
U^{\dagger} (\rlap/\partial U ) {\bf \Psi}_L
\label{1}
\ee
where $U$ is a chiral field, ${\bf \Psi}=({\bf \Psi}_R, {\bf \Psi}_L)$
is the nucleon field, and
$\rlap/\partial=\gamma^\mu \partial_\mu$.  In the low-energy
limit, the scattering amplitude given by (\ref{1}) is essentially
unique, given that the isospin of the nucleon is $\frac 12$
\cite{CCWZ}.

Ignoring isospin breaking and strong CP violation, the term which explicitly
breaks chiral symmetry must be a scalar-isoscalar. The simplest
non-trivial representation
of $SU (2)\times SU (2)$ which contains such a term is $(2,2)$, since
$(2,1)\oplus (1,2)$ contains only isospinors, and
$(1,3)\oplus (3,1)$ contains only isovectors. We therefore have,
\be
{\cal L}_2 = \frac 14\,f_{\pi}^2 m_{\pi}^2
\, {\rm Tr} ( U + U^{\dagger} )
-\frac {m_{\pi}^2}{\Lambda} \, {\overline {\bf \Psi}}{\bf \Psi}
\label{2}
\ee
We assume that $\Lambda$ is non-vanishing as $m_{\pi}\rightarrow 0$, so
that (\ref{2}) vanishes in the chiral limit. The nucleon mass is
$m_N = m_{\rm inv} + m_{\pi}^2/{\Lambda}$. The second term in (\ref{2}) is
usually dropped ($e.g.$ in chiral perturbation theory), but it is essential to
keeping the nucleons on shell and so we will retain it here.

{}From (\ref{1}-\ref{2}) it follows that the vector current is
\be
{\bf V}^a_{\mu} (x) =
&&i\frac{f_{\pi}^2}8 {\rm Tr} \left( [\tau^a , U^{\dagger} ]
\partial_{\mu} U \right) + h.c.
\nonumber\\
&&{}+\overline{\bf \Psi}\gamma_{\mu} \frac{\tau^a}2 {\bf \Psi} \nonumber\\
&&{}-\frac 14 (g_A-1)\overline{\bf \Psi}_L \gamma_{\mu} U^{\dagger}
[\tau^a , U ] {\bf \Psi}_L
\nonumber\\
&&{}+\frac 14 (g_A-1)\overline{\bf \Psi}_R \gamma_{\mu}
[\tau^a , U ] U^{\dagger} {\bf \Psi}_R
\label{3}
\ee
the axial current is
\be
{\bf A}^a_{\mu} (x) =
&&i\frac{f_{\pi}^2}8 {\rm Tr} \left( \left\{\tau^a , U^{\dagger} \right\}
\partial_{\mu} U\right) + h.c.
\nonumber\\
&&{}+\overline{\bf \Psi}\gamma_{\mu} \gamma_5\frac{\tau^a}2 {\bf \Psi}
\nonumber\\
&&{}-\frac 14 (g_A-1)\overline{\bf \Psi}_L \gamma_{\mu} U^{\dagger}
\left\{\tau^a , U \right\} {\bf \Psi}_L
\nonumber\\
&&{}+\frac 14 (g_A-1)\overline{\bf \Psi}_R \gamma_{\mu}
\left\{\tau^a , U \right\} U^{\dagger} {\bf \Psi}_R
\label{4}
\ee
and the scalar density is
\be
\sigma (x ) = \frac 1{m_{\pi}^2f_{\pi}} \,{\cal L}_2\, =
\frac{f_{\pi}}4 {\rm Tr}\left( U + U^{\dagger} \right)
-\frac 1{f_{\pi}\Lambda} \overline{\bf \Psi} {\bf \Psi}.
\label{5}
\ee

We also introduce the PCAC pion field
\be
\pi^a  (x ) = &&\frac 1{m_{\pi}^2 f_\pi} \partial^{\mu} {\bf A}_{\mu}^a
\nonumber\\ = &&
-i\frac{f_{\pi}}4 {\rm Tr}\left( \tau^a (U - U^{\dagger}) \right)
+\frac 1{f_{\pi}\Lambda} \overline{\bf \Psi} i\gamma_5\tau^a {\bf \Psi}
\label{6}
\ee
and the one-pion reduced axial current
\be
&&{\bf j}_{A\mu}^a = {\bf A}_{\mu}^a + f_{\pi} \partial_{\mu} {\pi}^a
\nonumber\\
= &&g_A\overline{\bf \Psi}\gamma_{\mu} \gamma_5\frac{\tau^a}2 {\bf \Psi}
+\frac 1{\Lambda} \partial_{\mu}
\left( \overline{\bf \Psi} i\gamma_5 \, \tau^a {\bf \Psi} \right)
+{\cal O} (\pi^3)
\label{7}
\ee

Between nucleon states of momentum $p_i$ and an implicit spin
dependence $s_i$,
\be
\lefteqn{\langle N(p_2) | {\bf A}^a_{\mu} (x) |N(p_1) \rangle  =
e^{i(p_2-p_1)\cdot x}}
\nonumber\\
& &\times\overline{u} (p_2) \left(\gamma_{\mu} \gamma_5 \,G_1 (t) +
(p_2-p_1)_{\mu} \gamma_5 \,G_2 (t) \right) \frac {\tau^a}2 \,\,u(p_1)
\label{8}
\ee
and
\be
\lefteqn{\langle N(p_2) | {\bf j}^a_{A\mu} (x) |N(p_1) \rangle  =
e^{i(p_2-p_1)\cdot x}}
\nonumber\\
& &\times\overline{u} (p_2) \left(\gamma_{\mu} \gamma_5 \,G_1 (t) +
(p_2-p_1)_{\mu} \gamma_5 \,\overline{G}_2 (t) \right)
\frac {\tau^a}2 \,\,u(p_1)
\label{9}
\ee
with $t=(p_2-p_1)^2$ and $\overline{G}_2$ is free of pion poles.

{}From (\ref{6}-\ref{7}), we also have $\partial^{\mu} {\bf j}_{A\mu}
= f_{\pi} (\Box + m_{\pi}^2 ) \, \pi$. Hence,
\be
&&\langle N(p_2) | \pi^a (x) |N(p_1) \rangle  =
\langle N(p_2) | \pi^a_{\rm in} (x) |N(p_1) \rangle \nonumber\\
&&\quad-\frac 1{f_{\pi}}
\int \,d^4y\, \Delta_R (x-y)
\langle N(p_2) | \partial^{\mu} {\bf j}^a_{A\mu} (y) |N(p_1) \rangle
\nonumber\\
&&= -\frac 1{f_{\pi}} \frac 1{t-m_{\pi}^2}
\overline{u} (p_2) \left( 2m_N\,G_1 (t) +
t\,\overline{G}_2 (t) \right)\nonumber\\&&\quad \qquad \times
i\gamma_5\,\frac {\tau^a}2 \,\,u(p_1)
e^{i(p_2-p_1)\cdot x}
\label{10}
\ee
where $\pi^a_{\rm in}$ is the incoming pion field, and we have used
$\langle N(p_2) | \pi^a_{\rm in} (x) |N(p_1) \rangle =0$. This is a
non-trivial requirement if the nucleon is a chiral soliton. This point
will not be pursued further here.

It follows from (\ref{7}-\ref{10}) that
\be
G_2(t) = \frac1{m_\pi^2-t}\left(2m_N G_1(t) + m_\pi^2
\overline{G}_2(t) \right).
\label{G2}
\ee
Also by definition, eq. (\ref{10}) is equal to
\be
-g_{\pi NN} (t) \frac 1{t-m_{\pi}^2} \overline{u} (p_2) i\gamma_5 \tau^a u
(p_1)\, e^{i (p_2-p_1)\cdot x}
\label{4.5}
\ee
and hence
\be
f_{\pi} g_{\pi NN} ( t )= m_N G_1 (t ) + \frac{t}2
\overline{G}_2 ( t )
\label{12}
\ee
where $g_{\pi NN} = g_{\pi NN} (m_{\pi}^2 )$ is the pion-nucleon coupling
constant. Extrapolating from $t=m_{\pi}^2$ to $t=0$ gives the standard
Goldberger-Treiman relation $g_A m_N \sim f_{\pi} g_{\pi NN}$, where $g_A=G_1
(0)$. However one can do better. Substituting (\ref{7}) at tree level
into (\ref{9}) gives
\be
&&G_1 (t) = g_A \nonumber\\
&&\overline{G}_2 (t) = -\frac{2}{\Lambda} .
\label{13}
\ee
Inserting (\ref{13}) into (\ref{12}) at $t=m_{\pi}^2$ gives the desired
relation
\be
\sigma_{\pi N} \equiv
\frac{m_{\pi}^2}{\Lambda} = g_A m_N - f_{\pi} g_{\pi NN}
\label{14}
\ee
between the pion-nucleon sigma term and the Goldberger-Treiman discrepancy.
The one-loop corrections to (\ref{15}) are of order ${m_N
m_{\pi}^2}/{(4\pi f_{\pi})^2}$. They will be discussed below.

Numerically, there is a huge cancellation in the right hand side, and
the value of $\sigma_{\pi N}$ is very sensitive to $g_{\pi
NN}$. Using the central values for all experimentally measured
quantities with $(m_N , f_{\pi}) = (940 , 92.4)$ MeV \cite{DATABOOK} and
$g_A= 1.2650(16)$ \cite{GA}, we have $\sigma_{\pi N} = -62$ MeV for the
value $g^2_{\pi NN}/4\pi =14.6(3)$ \cite{HIGH}, whereas we have
$\sigma_{\pi N} = 17$ MeV for the value $g^2_{\pi NN}/4\pi = 12.80(36)$
\cite{LOW}.

Unfortunately, this sensitivity means that we cannot directly extract
a reliable value of the sigma term from the existing data, although
the tree level result (\ref{14}) suggests a low value for the
pion-nucleon coupling constant, in view of the current value
$\sigma_{\pi N} = 45 \pm 8$ MeV \cite{SIGMACPT}.  We therefore turn to
the relation with Weinberg's formula for the pion-nucleon scattering
amplitude $i{\cal T}$ \cite{WEINBERG}. Taking $(k_1,a)$ as the
incoming pion, and $(k_2,b)$ as the outgoing pion, with
$p_1+k_1=p_2+k_2$, the formula reads
\be
i{\cal  T} =  i{\cal T}_V + i{\cal T}_S + i {\cal T}_{AA}
\label{15}
\ee
where
\be
i{\cal T}_V &=&
-\frac 1{f_{\pi}^2} k_1^{\mu}
\epsilon^{bac} \langle N(p_2) | {\bf V}_{\mu}^c (0)
| N(p_1) \rangle
\label{16}
\\
i{\cal T}_S &=&
-\frac i{f_{\pi}} m_{\pi}^2 \delta^{ab}
\langle N(p_2 ) | {\sigma} (0) | N(p_1) \rangle _{\rm conn.}
\label{17}
\\
i{\cal T}_{AA} &=&
-\frac1{f_{\pi}^2} k_1^\mu k_2^\nu
\int d^4x e^{-k_1\cdot x}\nonumber\\&&\times
\langle N(p_2) | T^*
{\bf j}_{A\mu}^a (x) {\bf j}_{A\nu}^b (0)| N(p_1)\rangle _{\rm conn.}\,\,.
\nonumber\\
\label{18}
\ee
Substitution of (\ref{3},\ref{5},\ref{7})
at tree level yields
\be
{\cal T}_V &=&
\frac 1{f_{\pi}^2} i \epsilon^{bac}
\overline{u} (p_2) \,\rlap/k_1\, \frac {\tau^c}2 u (p_1)
\label{19}
\\
{\cal T}_S &=&
\frac {m_{\pi}^2}{f_{\pi}^2\Lambda} \delta^{ab}
\;\overline{u} (p_2)  \,u (p_1)
\label{20}
\\
{\cal T}_{AA} &=&- \frac 1{f_{\pi}^2} \overline u (p_2)
\left(g_A \rlap/k_2 + \frac {2m_\pi^2}{\Lambda} \right)
\frac{\tau^b\tau^a}4
\nonumber\\
& &{}\times\frac 1{\rlap/{p}_1 +\rlap/{k}_1 +m_N}
\left(g_A \rlap/k_1 + \frac {2m_\pi^2}{\Lambda} \right) u(p_1)
\nonumber\\
&& +\left( k_1, a \leftrightarrow -k_2,b \right).
\label{21}
\ee
The isospin structure is decomposed as ${\cal T}^{ba}=\delta^{ab}
{\cal T}^+ + i\epsilon^{bac} \tau^c {\cal T}^-$ to give
\be
{\cal T}^+ = {\cal T}_S^+ + {\cal T}_{AA}^+
\qquad \qquad
{\cal T}^- = {\cal T}_V^- + {\cal T}_{AA}^-.
\label{22}
\ee
At threshold in the center of mass frame, the amplitudes ${\cal
T}^\pm$ can be extrapolated from data and written as scattering
lengths $a^\pm$. Taking (\ref{22}) at threshold,
\be
&&4\pi\left(1+\frac{m_\pi}{m_N}\right) a^+ = \frac{\sigma_{\pi
N}}{f_\pi^2} \left(1-\frac{\sigma_{\pi N}}{m_N} \right)
\nonumber\\
&&\hspace{1cm}-\frac1{f_\pi^2m_N} \frac{m_\pi^2}{4m_N^2-m_\pi^2}
\left(g_Am_N-\sigma_{\pi N}\right)^2
\label{23}
\ee
\be
&&4\pi\left(1+\frac{m_\pi}{m_N}\right) a^- =
\frac{m_\pi}{2f_\pi^2} \left(1-g_A^2\right)
\nonumber\\
&&\hspace{1cm}+ \frac{m_\pi}{f_\pi^2}
\frac{2}{4m_N^2-m_\pi^2} \left(g_Am_N-\sigma_{\pi N}\right)^2
\label{24}
\ee
showing the corrections to the Tomozawa-Weinberg formula are small.
Eqs. (\ref{14}) and (\ref{24}) give a direct relation between $a^-$ and
$g_{\pi NN}$, which is
\be
4\pi\left(1+\frac{m_\pi}{m_N}\right) a^- =
&&\frac{m_\pi}{2f_\pi^2} \left(1-g_A^2\right)\nonumber\\
&&+\frac{2m_{\pi}}{4m_N^2-m_\pi^2}g_{\pi NN}^2.
\label{25}
\ee
Using $a^-=(9.2\pm0.2)\times10^{-2}/m_\pi$
\cite{HELSINKI}, we find $g_{\pi NN}^2/4 \pi=14.4$. In terms of (\ref{24}),
eq. (\ref{23}) can be resolved into
\be
&&4\pi\left(1+\frac{m_\pi}{m_N}\right) \left( a^+  +\frac
{m_{\pi}}{2m_N} a^-\right)\nonumber\\=&&
\frac{\sigma_{\pi
N}}{f_\pi^2} \left(1-\frac{\sigma_{\pi N}}{m_N} \right)
+ \frac{m_{\pi}^2}{4f_{\pi}^2m_N} (1-g_A^2)\,\,.
\label{26}
\ee
Using $a^+=-(8\pm4)\times10^{-3}/m_\pi$ \cite{HELSINKI} and the above value
for $a^-$ gives $\sigma_{\pi N}= 2$ MeV. (The other root $\sigma_{\pi
N}\sim m_N$ has been discarded). In (\ref{14}), this corresponds to
the value $g_{\pi NN}^2/4\pi = 13.1$, to be compared with 14.4.

The present analysis can be extended to one-loop by using power
counting in $1/f_{\pi}$ \cite{YAMZAH1,YAMZAH2}. In this context we
have analyzed one-loop corrections to the above and they require a new
subtraction constant in $\overline{G}_2$. The extra piece of
data necessary to fix this constant is $g_p = m_{\mu} G_2
(-0.88m_{\mu}^2) = 8.2\pm 2.4$ available from muon capture in hydrogen
\cite{CAPTURE}. The loop corrections are in general small as can be
seen in Fig. 1 by the shift from the tree level (dotted line) to the
one-loop result at $\sigma_{\pi N}=0$. The exception is
$\overline{G}_2$ due to the large cancellation at tree level, since it
is proportional to $\sigma_{\pi N}$ in this case (eq. (\ref{13})). If we
require that $\sigma_{\pi N}$ is positive, the one-loop
correction does not exceed 50\%, and $g_{\pi NN}$ is larger than
the lower bound from \cite{LOW}, we then obtain an inequality between
$\sigma_{\pi N}$, $g_{\pi NN}$ and $g_p$, as indicated by the shaded
area of Fig. 1. We therefore have,
\be
12.4\leq \frac {g_{\pi NN}^2}{4\pi} \leq 13.15\qquad{\rm and}\qquad
8.30\leq g_p \leq 8.55
\label{27}
\ee
with $0\leq \sigma_{\pi N} \leq 70$ MeV, to one-loop. Our allowed
range for $g_P$ is to be compared with $8.44\pm0.16$ from
\cite{meiss}.

The justification of the supplementary rules, and details of the one-loop
calculation will be given elsewhere \cite{YAMZAH2}.

\begin{figure}
\epsfxsize=3.0in
\epsffile{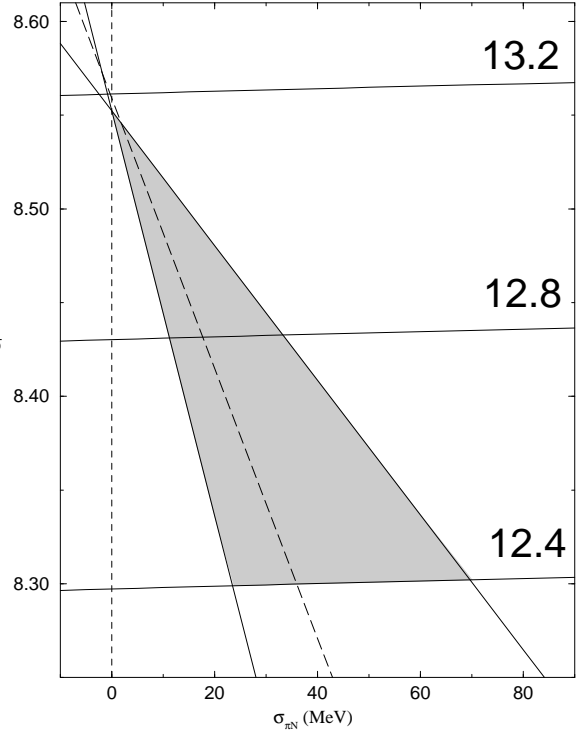}
\caption{The dependence of the pseudoscalar coupling constant ($g_P$)
on the $\pi N$-sigma term.  The horizontal lines have $g_{\pi
NN}^2/4\pi=12.4$, $12.8$, and $13.2$ respectively. The dotted line is
the tree result for $g_P$. Constraining the loop corrections of
$\overline{G}_2$ to be 50\% or less gives the shaded region. See the
text for further discussion.}
\end{figure}

\vglue 0.6cm
{\bf \noindent  Acknowledgements \hfil}
\vglue 0.4cm
This work was supported in part  by the US DOE grant DE-FG-88ER40388.

\vskip 1cm
\setlength{\baselineskip}{15pt}

\end{document}